\documentclass[11pt,a4paper]{article}
\usepackage{amsmath,amsfonts,amssymb,amsthm,cite}
\evensidemargin=\oddsidemargin
\advance\topmargin by -\headheight
\advance\topmargin by -\headsep
\numberwithin{equation}{section}           

\newcommand{\bx}{\bar{x}}                  
\newcommand{\by}{\bar{y}}
\newcommand{\bz}{\bar{z}}
\newcommand{\tu}{\tilde{u}}
\newcommand{\tv}{\tilde{v}}
\newcommand{\tx}{\tilde{x}}                
\newcommand{\ty}{\tilde{y}}
\newcommand{\tz}{\tilde{z}}

\renewcommand{\a}{\alpha}                    
\renewcommand{\b}{\beta}                     
\newcommand{\dl}{\delta}                     
\newcommand{\tri}{\Delta}                    
\newcommand{\ep}{\epsilon}                   
\newcommand{\eps}{\varepsilon}               
\newcommand{\vf}{\varphi}                    
\newcommand{\ga}{\gamma}                     
\newcommand{\Ga}{\Gamma}                     
\newcommand{\la}{\lambda}                    
\def\m{\mu}                                  
\def\n{\nu}                                  
\renewcommand{\th}{\theta}                   
\newcommand{\Th}{\Theta}                     
\newcommand{\pa}{\partial}                   
\newcommand{\F}{\mathcal{F}}                 
\renewcommand{\SS}{\mathcal{S}}              
\newcommand{\Sf}{\mathbb{S}}                 
\newcommand{\half}{\tfrac{1}{2}}             
\newcommand{\ihalf}{\tfrac{i}{2}}            
\newcommand{\fourth}{\tfrac{1}{4}}           
\newcommand{\Z}{\mathbb{Z}}                  
\newcommand{\R}{\mathbb{R}}                  
\newcommand{\T}{\mathbb{T}}                  
\newcommand{\x}{\times}                      
\newcommand{\sepword}[1]{\quad\mbox{#1}\quad} 
\DeclareMathOperator{\Tr}{Tr}                

\def\ds{\displaystyle}
\def\sc{\scriptstyle}
\def\igual{\!\!\!\!=\!\!\!}
\def\mas{\!\!\!\!+\!\!\!}
\def\menorigual{\!\!\!\!\leq\!\!\!}

\begin{document}
\thispagestyle{empty}

\begin{titlepage}
\rightline{CPT-2005/P.014}
\rightline{UCM-FTI-05/121}

\vskip 40pt
\centerline{\Large\bf{Position-dependent noncommutative products:}}
\vskip 6pt
\centerline{\Large\bf{classical construction and field theory}}
\vskip 40pt
\centerline{\Large V. Gayral${}^{a,}$\footnote{Also at Universit\'e de
               Provence, gayral@cpt.univ-mrs.fr},
               J.~M. Gracia-Bond\'{\i}a${}^b$ and F. Ruiz Ruiz${}^b$}
\vskip 10pt
\begin{center}
{\it ${}^a$ Centre de Physique Th\'eorique, UMR 6207, 13288
        Marseille, France.\\
      ${}^b$ Departamento de F\'{\i}sica Te\'orica I,
        Universidad Complutense de Madrid, 28040 Madrid, Spain.}
\end{center}

\vspace{60pt}

\centerline{\bf Abstract}
\medskip
{\leftskip=40pt \rightskip=40pt
We look in Euclidean $\R^4$ for associative star products realizing
the commutation relation $[x^\mu,x^\nu]=i\Theta^{\mu\nu}(x)$, where
the noncommutativity parameters $\Theta^{\mu\nu}$ depend on the
position coordinates $x$. We do this by adopting Rieffel's deformation
theory (originally formulated for constant $\Th$ and which includes
the Moyal product as a particular case) and find that, for a topology
$\R^2\times\R^2$, there is only one class of such products which are
associative. It corresponds to a noncommutativity matrix whose
canonical form has components $\Th^{12}=-\Th^{21}=0$ and
$\Theta^{34}=-\Theta^{43}= \theta(x^1,x^2)$, with $\th(x^1,x^2)$ an
arbitrary positive smooth bounded function. In Minkowski space-time,
this describes a position-dependent space-like or magnetic
noncommutativity. We show how to generalize our construction to~$n\geq
3$ arbitrary dimensions and use it to find traveling noncommutative
lumps generalizing noncommutative solitons discussed in the
literature. Next we consider Euclidean $\la\phi^4$ field theory on
such a noncommutative background. Using a zeta-like regulator, the
covariant perturbation method and working in configuration space, we
explicitly compute the UV~singularities. We find that, while the
two-point UV~divergences are non-local, the four-point UV~divergences
are local, in accordance with recent results for constant $\Theta$.

\vskip60 pt\noindent
{\it PACS:\/} 11.10.-z, 11.10.Lm, 11.10.Gh \\
{\it Keywords:} Noncommutative field theory;
Non-local models; Renormalization; Generalized deformations\\
\par }

\end{titlepage}


\newpage
\setcounter{page}{2}
\section{Introduction.}
\label{sec:introibo}

Seiberg and Witten have argued~\cite{SeibergW} that the endpoints of
open strings in a magnetic field background live on a Moyal
(hyper)plane. The mathematical properties of these spaces were
familiar to physicists since the fifties, especially in connection
with the phase-space formulation of quantum mechanics~\cite{Moyal}. To
this knowledge must be added the recent proof~\cite{Himalia} that
Moyal planes are spectral triples, that is, virtual spin manifolds in
the sense of Connes~\cite{ConnesGrav,Polaris}. Yet, despite all
these good properties, quantum field theory on Moyal planes (NCFT for
short) is a difficult construct.

The problems of NCFT arise with renormalization and are ultimately
caused by the non-local nature of the Moyal `star' product. One of
these problems is the conciliation of unitarity with the locality of
the counterterms needed to subtract UV~divergences. Indeed, whereas on
the one hand it has been proved~\cite{GomisM} that noncommutativity
matrices $\Theta=[\Theta^{\m\n}]$ with only space-space components
yield one-loop renormalized Green functions consistent with unitarity,
on the other hand it has been shown that such Green functions do not
define an effective action~\cite{GGBRR} (see ref.~\cite{Bahns} for an
argument in terms of Wick products).  This indicates that
noncommutativity matrices with nonvanishing space-time components
should be investigated. Some progress along this line is taking
place~\cite{Bahnsetalt,Fujikawa}. Another source of trouble is the
occurrence of tachyonic instabilities~\cite{tachyon,FRRsusy} already
at first order in perturbation theory. Such instabilities, and partly
the non-locality of the UV counterterms for space-space
matrices~$\Theta$, are linked to the by now well-known UV/IR mixing
phenomenon~\cite{Minwallaetal}. Fortunately enough, there is a cure
for this problem, since such instabilities can be eliminated and the
UV/IR mixing avoided by introducing
supersymmetry~\cite{Matusis,FRRsusy}. So, all in all, progress is
being made little by little. In this respect it is worth mentioning
the general belief that by removing the UV/IR mixing the
renormalization program is feasible at higher orders in perturbation
theory.

In this paper we take a different view and investigate almost
unexplored territory: NCFT on non-constant (i.e. coordinate-dependent)
noncommutativity spaces.  We have several motivations for this
enterprise.  The first one comes from the appearance of non-constant
noncommutativity in various contexts, in particular from the continued
development of string theory; a literature sample may include
refs.~\cite{Selene,Lizzi,Dolan,CalmetW,BehrS,Hashimoto}.  The second
one is that, partly motivated by the former and by the hope of dealing
with gravity, it is natural to look at the noncommutativity parameters
as dynamical variables.  A step in this direction is to make them
dependent on coordinates.  One could count, in the third place, the
hope of finding a transitional regime between the noncommutative and
the commutative realms.  This hope seems to be misplaced, at least for
non-supersymmetric theories, as the UV/IR phenomenon and the
generation of IR singularities when the noncommutativity parameters
approach zero are not overcome.  Last, but not least, the problem has
interest from the viewpoint of the general theory of noncommutative
spaces, for there are not so many examples of these.

As we will see below, there are ways to realize the commutator
\begin{equation}
    [x^\m,x^\n] = i\Th^{\m\n}(x)\,,
\label{granted}
\end{equation}
where the matrix $\Th(x)=[\Th^{\m\n}(x)]$ depends on coordinates, by
means of star products.  Nonassociative star products have been
explicitly used in~\cite{Cornalba}. However, for field theory and also
for the sake of defining noncommutative spaces with a minimum set of
bona fide properties, one wants an \textit{associative} product with
more or less the same basic properties as Moyal's.  So we must first
investigate the question of whether there are admissible associative
generalizations of the Moyal product with position-dependent
noncommutativity.  We will do this by using Rieffel's deformation
procedure~\cite{RieffelDefQ}, a true and tested recipe for generating
suitable noncommutative spaces with constant noncommutativity, and
studying under what conditions it can be extended to cover
non-constant noncommutativity.  This by itself is not a trivial issue.
In fact, in the literature one often finds eq.~(\ref{granted}), but no
direct explicit realizations are given.  As already declared, we want
to do quantum field theory on $\R^4$ with non-constant $\Th(x)$, so we
will be interested in non-periodic star products of functions (fields)
defined on $\R^4$.  We will consider Euclidean~$\R^4$ with topology
$\R^2\times\R^2$ and use coordinates $x=(\bx^1,\bx^2,\tx^1,\tx^2)$ to
exploit the canonical form of antisymmetric matrices $\Th(x)$ in four
dimensions and the fact that $\Th(x)$ can only have rank~4, 2 or 0.
The outcome of this investigation is surprising because of the
uniqueness of its result.  Namely, the only noncommutativity matrix
$\Th(x)$ yielding an associative non-constant star product of the
desired type will turn out to have rank~2 and
entries~$\Th^{\bar{i}\,\!\bar{j}}\!=0$ and
$\Th^{\tilde{a}\tilde{b}}=\eps^{\tilde{a}\tilde{b}}\th(\bx)$, with
$\th(\bx)$ an arbitrary positive and suitably bounded function, or the
other way around.  This corresponds in the Minkowskian framework to an
$x$-dependent generalization of space-like or magnetic constant
non-commutativity.  We will see all this in Section~2.  We also give
there an example of a non-constant star product for $\R^4$ foliated by
3-spheres.

The construction in Section 2 of a position-dependent star product
makes an obvious exercise to generalize the noncommutative soliton
solutions in the literature~\cite{Gopakumar} to traveling
noncommutative lumps. This is done in Section 3.

Once we have a sensible non-constant star product, we move on to
consider quantum $\la\phi^4$ field theory for such a product. We will
define the quantum theory through the effective action and study
whether the theory is renormalizable at one loop. Because of the
presence of the generic function~$\th(\bx)$ we are led to work in
position space. To compute one-loop radiative corrections and examine
their UV singularities we will use the covariant perturbation
method~\cite{covariant}, due to Barvinsky and Vilkovisky, proposed as
an alternative to the Schwinger--DeWitt technique~\cite{Sch--DeWitt}
and tested in the quantization of fields coupled to gravitational
backgrounds. We discuss in Section~4 the particulars of the method
when applied to our problem.

Sections~5 and~6 contain the analysis of the two and four-point parts
of the effective action. Despite the non-localities of the theory, we
succeed in completely characterizing the UV~divergences in the model.
While the four-point divergences are completely local and can be
subtracted by local counterterms, the two-point divergences are
non-local and cannot be removed by local counterterms. This poses an
obstacle to traditional perturbative renormalization and shows that
the $\la\phi^4$ model can only be viewed at this stage as an effective
theory. This is in complete accordance with the $\th$-constant case.
Section~7 contains our conclusion. We include at then end two
mathematical appendices~~---which the not mathematically inclined
reader may omit---~~to put our discussion on a rigorous footing.

We want to emphasize at the outset that we are not claiming ours
is the only method to construct associative star products compatible
with position-dependent commutation relations.  Other methods are
found in the literature, for instance in relation with Kontsevich's
formality for Poisson structures~\cite{Kon}, or as deformations of the
commutation relation~\eqref{granted} for constant $\Th$
\cite{FoscoT}. Those methods have in common that the noncommutativity
matrix is regarded as a perturbation, and the expansions in~$\Th$ are
analytically uncontrolled. For the noncommutative products obtained in
ref.~\cite{Selene} as symplectic reductions of Moyal algebras to
function groups, this is also true in practice, although not in
principle. Our approach, however, is not perturbative in $\Th$. As we
intend here to do field theory, well-definiteness of the various
objects involved in our analysis becomes crucial, so our preferences
go to deformations with good analytical properties, and in that
respect Rieffel's approach singles itself.

\section{Construction of star products with non-constant
   $\Th(x)$.}

We start by recalling the construction of the original Moyal product.
For any finite dimension $k\ge2$, we denote by $\Theta$ a real
non-degenerate skew-symmetric $k \x k$ matrix. The non-degeneracy
condition implies even dimension, so that $k = 2N$ with $N\ge 1$.
Given now two suitable functions $f$ and $g$ on $\R^{2N}$, their Moyal
product is a function of the same type defined as
\begin{equation}
    (f \!\star_\Th\! g)\,(x) = \frac{1}{\pi^{2N}\det\Theta}
      \int d^{2N}\!y \, d^{2N}\!z \, f(x + y)\, g(x + z)\,
           e^{-2iy\Theta^{-1}z}\,.
\label{eq:Moyal-prod}
\end{equation}
Since $y$ and $z$ are points in space-time, the entries of~$\Theta$
have the dimensions of an area. The Moyal product clearly satisfies
the trace property, for
\begin{equation*}
    \int d^{2N}\!x \,(f\!\star_\Theta\! g) \,(x) =
        \int d^{2N}\!x\, f(x)\,g(x)\,.
\end{equation*}
This property allows to extend definition~\eqref{eq:Moyal-prod} to
(huge) appropriately chosen function and distribution multiplier
spaces~\cite{PhobosandDeimos}. In particular, the Moyal product of
periodic functions is periodic, and thus noncommutative tori are
subsumed in Moyal theory.

The crucial remark by Rieffel is that eq.~\eqref{eq:Moyal-prod} may be
rewritten as
\begin{equation}
    (f \!\star_\Theta\! g) \,(x) = \frac{1}{(2\pi)^k} \int d^k\!y
       \, d^k\!z\, f(x - \half\Theta y)\, g(x + z)\, e^{-iyz}\,.
\label{eq:Moyal-prod-rewritten}
\end{equation}
This formula is the starting point for a far-reaching deformation
theory of algebras, in which the~$\Theta$-parameters carry actions of
continuous abelian groups~$\R^m\x\T^{l-m}$ and no longer
non-degeneracy/even dimension are issues. For instance, a
generalization of eq.~\eqref{eq:Moyal-prod-rewritten} in terms of such
actions, by isometries on suitable Riemannian manifolds, has been
analyzed in refs.~\cite{Gayral,Himaliatwo}, with most properties
of the Moyal product being kept.

To begin our study, we recall the well-known fact that any
antisymmetric constant matrix $\Theta$ can be written in the form
\begin{equation*}
    \Theta = A^t\begin{pmatrix} \zeta\,S & 0 \\ 
                                0 & \theta\,S \end{pmatrix} A 
    \sepword{with} S = \begin{pmatrix} 0 & 1 \\ -1 & 0 \end{pmatrix}\,,
\end{equation*}
where $A$ is orthogonal, $A^t=A^{-1}$, and $\zeta$ and $\th$ are
constants. Note that, being $A$ orthogonal, the previous
transformation is both a congruence and a similarity. Let us assume
now that two antisymmetric matrices $\Th$ and $\Th'$ are related by
such a transformation,
\begin{equation*}
\Theta' = A^t\Theta A =: {\cal A}[\Theta],
\end{equation*}
and let us define local transformations
\begin{equation*}
   Af(x) = f(A^{-1}x).
\end{equation*}
As is well known, it then follows that the corresponding star products
$\star_\Th$ and $\star_{\Th'}$ are covariant under $A$, meaning that
there exists an algebra isomorphism such that, with a slight abuse of
notation,
\begin{equation}
    A\left(f\!\star_{{\cal A}[\Th]}\! g\right)\!(x) 
        = Af(x)\!\star_{\Theta}\!Ag(x)\,.
\label{constant-cov}
\end{equation}
In particular, if $\Theta$ and~$A$ commute, there is
equivariance, i.e.
\begin{equation*}
    A(f\!\star_{\Theta}\! g)(x) = Af(x)\!\star_{\Theta}\! Ag(x)\,.
\end{equation*}
For instance, if 
\begin{equation*}
  \Theta=\begin{pmatrix} \zeta\,S & 0 \\ 0 & \theta\,S
  \end{pmatrix}\,, 
\end{equation*}
the star product is equivariant under transformations of the group
$SO(2;\R)\x SO(2;\R)$ ---actually under the bigger group $SL(2;\R)\x
SL(2;\R)$ of symplectic transformations.

For $x$-dependent $\Th$ we adopt Rieffel's formula
\eqref{eq:Moyal-prod-rewritten} as our starting point. In this case it
is natural to consider a large `gauge' group $\F(\R^4,O(4;\R))$ of
transformations
\begin{equation*}
   A(x)f(x) = f(A^{-1}(x)\,x)
\end{equation*}
for suitably well-behaved orthogonal-matrix valued functions $A(x)$.
Then it is not difficult to see that, instead of
eq.~\eqref{constant-cov}, with $A^t(x)\Theta(y)A(x) =: {\cal
A}(x)[\Theta(y)]$, the covariance property reads:
\begin{equation*}
   A(x)\!\left(f\!\star_{{\cal A}(x)[\Th(A(x)x)]}\! g\right)\!(x) 
       = A(x)f(x)\!\star_{\Th(x)}\! A(x)g(x)\,.
\end{equation*}
Thus the product $\star_{{\cal A}(x)[\Th(A(x)x)]}$ is associative if
and only if $\star_{\Th(x)}$ is. This observation allows us to reduce
the study of associativity of~$\star_{\Th(x)}$ to the case in which
the matrix $\Th(x)$ is brought into the canonical form
\begin{equation*}
     \Th(x) = \begin{pmatrix} \zeta(x)\,S & 0 \\
                              0  & \theta(x)\,S
              \end{pmatrix}
     \sepword{with} S = \begin{pmatrix} 0 & 1 \\ -1 & 0 \end{pmatrix}\,,
\end{equation*}
$\zeta(x)$ and $\theta(x)$ being functions of $x$. Henceforth we
suppose that this reduction has taken place. Recalling that for a
point~$x\in\R^4$ we are writing $x=(\bx,\tx)=(\bx^i,\tx^a)$, with
$i,a=1,2$, the latter defines the star product of two functions $f$
and~$g$ on~$\R^4$ as
\begin{equation}
   (f \!\star_\Th\! g) (x)
     = \frac{1}{(2\pi)^4} \!\int d^4\!y\,d^4\!z~
         f\big(\bx-\half\zeta(x)S\by,~\tx-\half\theta(x)S\ty\big)~
          g(\bx+\bz,\,\tx+\tz)~e^{-i(\by\bz+\ty\tz)}\,.
\label{eq:deformation}
\end{equation}
For this definition to be pertinent, we need two conditions. First, it
must realize the commutation relations
\begin{equation}
   [\bar{x}^i,\bar{x}^j] = i\Theta^{ij}(x) = i\eps^{ij}\,\zeta(x)
   \qquad
   [\tilde{x}^a,\tilde{x}^b] = i\Theta^{ab}(x) = i\eps^{ab}\,\theta(x)
   \qquad
   [\bar{x}^i,\tilde{x}^a]=0\,,
\label{cr}
\end{equation}
where $\eps^{12}=1$. Secondly, it must satisfy the basic properties of
the ordinary $\Theta$-constant Moyal product, associativity among
them. In this regard it is worth emphasizing that the entanglement
caused by a nonconstant $\Th$ in~\eqref{eq:deformation} threatens
associativity.

In this section we study whether the commutation relations are
realized and find the constraints that associativity imposes on the
functions $\zeta(x)$ and $\theta(x)$.  Clearly the matrix $\Th(x)$
will have rank~4, 2 or 0.  For analytical reasons discussed later, we
take the functions $\zeta(x)$ and $\theta(x)$ either zero or positive
everywhere. This implies that we restrict ourselves to constant
rank. Rank~4 corresponds to both functions $\zeta(x)$ and $\theta(x)$
being different from zero for all $x$, rank~2 corresponds to the
product $\zeta(x)\th(x)$ vanishing for all $x$ and rank~0 to both
being zero.  We do not consider the rank~0 case since it corresponds
to the ordinary commutative product. Let us consider the rank~4 and
rank~2 cases separately.

\subsection{Case 1: $\mathbf{rank}(\Theta)$ = 4.}

Using eq.~\eqref{eq:deformation} for the product
$\bx^i\!\star_\Theta\!\bx^j$ we have
\begin{equation*}
    \bx^i\!\star_\Theta\!\bx^j = \frac{1}{(2\pi)^4} \int
       d^2\!\by\,\,d^2\!\ty\,\,d^2\!\bz\,\,d^2\!\tz\,\,
       \big[\,\bx^i - \half\,\zeta(x)\,\eps^{ik}\,\by^k\big]\,
       \big[\bx^j + \bz^j\big] \, e^{-i(\by\bz+\ty\tz)}\,.
\end{equation*}
Expanding the product in the integrand and integrating over $\ty$ and
$\tz$ we have
\begin{equation*}
     \bx^i\!\star_\Theta\!\bx^j = \frac{1}{(2\pi)^2} \int
        d^2\!\by\,\,d^2\!\bz\,\,  \Big[\,\bx^i\bx^j + \bx^i\bz^j
           - \frac{1}{2}\,\zeta(x)\,\eps^{ik}\,\by^k\bx^j
           - \frac{i}{2}\,\zeta(x)\,\eps^{ik}\,\by^k\,
            \frac{\pa}{\pa\by^j}\Big]\, e^{-i\by\bz} \,.
\end{equation*}
Integrating now over $\by$ and $\bz$ we obtain the ordinary
product $\bx^i\bx^j$ for the first term, zero for the second and
third terms and $\half\zeta(x)\eps^{ij}$ for the fourth term. Hence
\begin{equation*}
   \bx^i\!\star_\Theta\!\bx^j = \bx^i\bx^j
                     + \frac{i}{2}\,\zeta(x)\,\eps^{ij}\,.
\end{equation*}
It then follows that the generalized Rieffel
formula~\eqref{eq:deformation} realizes the first commutation relation
in eq.~(\ref{cr}). Analogous arguments show that the other two
commutation relations in~(\ref{cr}) are also implemented by
definition~\eqref{eq:deformation}.

We next discuss associativity,
\begin{equation}
   (f \star_\Theta g)\, \star_\Theta h
        = f \star_\Theta (g \star_\Theta h)\,.
\label{associativity}
\end{equation}
As is well known, a necessary condition, though not sufficient, for
associativity is the Jacobi identity
\begin{equation}
    [[f,g]_{\star_\Th},h]_{\star_\Th} +
    [[h,f]_{\star_\Th},g]_{\star_\Th} +
    [[g,h]_{\star_\Th},f]_{\star_\Th} =0 \,,
\label{Jacobi}
\end{equation}
where we have written $[f,g]_{\star_\Th}=f\star_\Th g - g\star_\Th
f$. Let us study whether it is fulfilled, prior to moving on to
associativity. For $f=\bx^i$, $g=\bx^j$ and $h=\bx^k$, since the
indices $i,j,k$ may only take values in the range 1,2, two of them
must be equal, say $i=j$. Then the identity~(\ref{Jacobi}) becomes
\begin{equation*}
  [[\bx^k,\bx^i]_{\star_\Th},\bx^i]_{\star_\Th} 
    +[[\bx^i,\bx^k]_{\star_\Th},\bx^i]_{\star_\Th}=0\,,
\end{equation*}
which is trivially satisfied. The same argument shows that the Jacobi
identity also holds for $f=\tx^a$, $g=\tx^b$ and $h=\tx^c$. Now, for
$f=\bx^i$, $g=\bx^j$ and $h=\tx^a$, the commutation
relations~(\ref{cr}) give for eq.~(\ref{Jacobi})
\begin{equation*}
[\zeta(x),\tx^a]_{\star_\Th} =0\,.
\end{equation*}
As we are assuming $\th(x)\ne0$, this requires the function $\zeta(x)$
not to depend on~$\tx$. Analogous arguments show that the Jacobi
identity for $f=\tx^a,g=\tx^b$ and $h=\tx^i$ demands the function
$\theta(x)$ not to depend on~$\bx$. We then have $\zeta(x)=\zeta(\bx)$
and $\theta(x)=\theta(\tx)$. In other, words the noncommutativity matrix
is the direct sum of two $2\x 2$ matrices that do not speak to each
other. 

Let us now consider associativity itself. Using~\eqref{eq:deformation}
and similar arguments as for the study of the commutation relations,
it is easy to see that
\begin{equation*}
   (\bx^i\! \star_\Theta \bx^j ) \star_\Theta \bx^k
     - \bx^i\! \star_\Theta (\bx^j \!\star_\Theta \bx^k)
   =-\, \frac{1}{4}~\zeta(\bx)\,\bigg[
       \eps^{ij}\eps^{kr}\,\frac{\pa\zeta}{\pa\bx^r}
     + \eps^{jk}\eps^{ir}\,\frac{\pa\zeta}{\pa\bx^r}\,\bigg]\,.
\end{equation*}
Associativity for $i\!=\!j\!=\!1,~k\!=\!2$ and for
$i\!=\!1,~j\!=\!k\!=\!2$ requires $\,\pa\zeta/\pa\bx^2\!=0\!\,$
and $\,\pa\zeta/\pa\bx^1\!=\!0$. Hence $\zeta(\bx)$ must be
constant. The same arguments applied to $\tx^a,\,\tx^b$ and $\tx^c$
show that $\theta(\tx)$ must also be constant. Therefore
\begin{equation*}
    {\rm rank~4:} \qquad \zeta(x)= {\rm const.}
                  \qquad \theta(x) = {\rm const.}
\end{equation*}
This brings us back to the ordinary $\Theta$-constant Moyal
4-dimensional star product. We thus conclude that for rank~4,
Rieffel's deformation framework does not provide generalizations to
non-constant star products in terms of associative algebras of
functions. From our arguments it trivially follows that there is no
2-dimensional generalization in the sense of Rieffel of the Moyal
product to non-constant $\Theta(x)$, either.

\subsection{Case 2: $\mathbf{rank}(\Theta)$ = 2.}

Constant rank~2 corresponds to one of the two functions $\zeta(x)$ and
$\th(x)$ being identically zero. So let us take $\zeta(x)=0$; if
$\theta(x)=0$ one proceeds similarly. In this case integration over
$\by$ and $\bz$ in eq.~\eqref{eq:deformation} is trivial and thus we
are left with
\begin{equation}
    (f \star_\theta g) (x) = \frac{1}{(2\pi)^2} \int
       d^2\!\tu\,d^2\!\tv\,\,f\big(\bx,\tx-\half\,\theta(x)\,S\tu\big)~
         g(\bx,\tx+\tv)\,\,e^{-i\tu\tv}\,.
\label{rank2}
\end{equation}
Here we have changed the notation to $\star_\theta$ to make explicit
that we are in the case of rank~2. The same arguments as for the
rank~4 case show that eq.~(\ref{rank2}) realizes the commutation
relations (\ref{cr}). The Jacobi identity can also be analyzed in the
same manner as for rank~4. The difference is that now it does not
restrict the dependence of $\theta(x)$ on $x$. Associativity, however,
does, for we now arrive at
\begin{equation*}
   (\tx^a\! \star_\theta \tx^b ) \star_\theta \tx^c
      - \tx^a\! \star_\theta (\tx^b \!\star_\theta \tx^c)
   = - \,\frac{1}{4}~\theta(x)\,\bigg[
       \eps^{ab}\eps^{cd}\,\frac{\pa\theta}{\pa\tx^d}
     + \eps^{bc}\eps^{ad}\,\frac{\pa\theta}{\pa\tx^d}\,\bigg]
\end{equation*}
and for the right-hand side to vanish for all $a,b,c=1,2$ we need
$\pa\theta/\pa\tx^1=0$ and $\pa\theta/\pa\tx^2=0$. Hence $\theta(x)$
may only depend on $\bx$,
\begin{equation*}
      {\rm rank~2:} \qquad \theta(x)=\theta(\bx)\,.
\end{equation*}

Once we have this, we prove associativity for arbitrary functions
$f,~g$ and $h$ on $\R^4$. For the left-hand side in
eq.~(\ref{associativity}) we have
\begin{align*}
  [(f \star_\theta g)\, \star_\theta h]\,(x) &= \frac{1}{(2\pi)^4}
    \int d^2\!\tu_1\,\,d^2\!\tv_1\,d^2\!\tu_2\,\,d^2\!\tv_2~
               e^{-i(\tu_1\tv_1+\tu_2\tv_2)}\\[1pt]
    &~ {\sc \times}\,
    f\big(\bx,~\tx-\half\,\theta(\bx)\,S(\tu_1+\tu_2)\big)~
    g\big(\bx,~\tx-\half\,\theta(\bx)\,S\tu_2+\tv_1\big)~
    h(\bx,\tx+\tv_2)~.
\end{align*}
Making the changes $\tv_1\to\tv_1+\half\,\theta(\bx)\,S\tu_2$ and
$\tu_2\to\tu_2-\tu_1$, we obtain after a plane wave integration
\begin{equation}
     \frac{1}{(2\pi)^2} \int d^2\!\tu\,d^2\!\tv~
       f\big(\bx,~\tx-\half\,\th(\bx)\,S\tu\big)~ g(\bx,\tx+\tv)
       ~h\big(\bx,~\tx+\tv+\half\,\th(\bx)\,S\tu\big)~e^{-i\tu\tv}.
\label{intermediate}
\end{equation}
For the other side we have
\begin{align*}
  [f \star_\theta\,( g \star_\theta h)]\,(x) &= \frac{1}{(2\pi)^4}
    \int d^2\!\tu_1\,\,d^2\!\tv_1\,d^2\!\tu_2\,\,d^2\!\tv_2~
               e^{-i(\tu_1\tv_1+\tu_2\tv_2)}\\[1pt]
    &~{\sc \times}\,f\big(\bx,~\tx-\half\,\theta(\bx)\,S\tu_2\big)~
     g\big(\bx,~\tx-\half\,\theta(\bx)\,S\tu_1+\tv_2\big)~
     h(\bx,\,\tx+\tv_1+\tv_2)\,.
\end{align*}
Performing the changes $\tv_2\to\tv_2-\tv_1$ and $\tv_1\to
\tv_1-\half\,\theta(\bx)\,S\tu_1$ in this order, and integrating over
$\tu_1$ and $\tv_1$, we obtain
\begin{equation*}
    \frac{1}{(2\pi)^2} \int d^2\!\tu\,d^2\!\tv~
       f\big(\bx,~\tx-\half\,\th(\bx)\,S\tu\big)~
         g\big(\bx,~\tx+\tv-\half\,\theta(\bx)\,S\tu\big)~
           h(\bx,\,\tx+\tv)~ e^{-i\tu\tv}\,.
\end{equation*}
Changing now $\tv\to\tv+\half\,\theta(\bx)\,S\tu$ we
reproduce~(\ref{intermediate}), which proves associativity.

Appendix A contains a discussion of other basic properties of this
star product. Among them, it is convenient to mention here the trace
property and the Leibniz rule.  The trace property
\begin{equation*}
    \int d^2\!\tx\,d^2\!\bx\, \big( f\star_\th g)(\bx,\tx)
        = \int d^2\!\tx\,d^2\!\bx\, f(\bx,\tx)\, g(\bx,\tx)
\end{equation*}
follows from eq.~(\ref{rank2}) and some simple changes of variables.
As concerns the Leibniz rule, differentials with respect to
$\tx$ coordinates satisfy it, for
\begin{equation}
     \frac{\pa}{\pa\tx^a}\, \big(f\star_\th g)
         = \frac{\pa f}{\pa\tx^a}\star_\th g
         + f \star_\th \frac{\pa g}{\pa\tx^a}\,,
\label{LeibnizOK}
\end{equation}
whereas derivatives with respect to $\bx$ coordinates do not. Indeed,
acting with $\pa/\pa\bx_j$ on eq.~(\ref{rank2}), recalling that
$\theta$ only depends on $\bx$ and integrating by parts, we have
\begin{equation}
    \frac{\pa}{\pa\bx^j}\, \big(f\!\star_\th g)
       = \frac{\pa f}{\pa\bx^j}\star_\th g
       + f \star_\th \frac{\pa g}{\pa\bx^j}
       + \frac{i}{2}\,\,\frac{\pa\th}{\pa\bx^j}~\eps^{ab}\,
         \Big( \frac{\pa f}{\pa\tx^a}\star_\th
                              \frac{\pa g}{\pa\tx^b}\Big) ~.
\label{Leibnizviol}
\end{equation}
As a consequence, the star product constructed here is not an
isospectral deformation in the sense of Connes and
Landi~\cite{ConnesLa,ConnesDV}, since commutators of the ordinary
Dirac operator on~$\R^4$ with left or right star multiplication
contain derivatives. In particular, the commutators are no longer
bounded operators. The formulation of non-constant noncommutativity
spaces as Connes' spectral triples is likely to involve techniques of
the noncommutative geometry of foliations~\cite{CM}.

The star product associated to $\theta(x)=\theta(\bx)$ and
$\zeta(x)=0$ that we have just constructed can be viewed as follows.
Suppose we take infinitely many copies of $\R^2$ and that on each copy
we define an ordinary $\theta$-constant Moyal product, with $\theta$
being different for different copies. To account for this difference,
we make $\th$ to depend on two real arbitrary parameters $\bx^1$ and
$\bx^2$, external to $\R^2$. Now we form the product of the deformed
$\R^2$, which we denote $\R^2_{\theta(\bx)}$, with the range of the
external parameters, which is nothing but $\R^2_{\bx}$. This gives a
foliation $\R^2_{\bx}\rtimes\R^2_{\theta(\bx)}$ of $\R^4$, i.e. the
structure we have. From this viewpoint, it is natural that this
construction define a non-constant $\Theta(x)$ star product. This is
the only one that happens to be associative within the framework
provided by Rieffel's deformation quantization method. The
construction generalizes to other dimensions in different forms.
Indeed, for~$n\geq 3$ odd or even and $n-2m\geq 1$, we can write
$\R^n\simeq\R^{n-2m}\rtimes\R^{2m}$ and $x=
(\bx^1,\ldots,\bx^{n-2m},\tx^1,\ldots,\tx^{2m})$. We can even
substitute a manifold for~$\R^{n-2m}$. The arguments presented here
all go through, thus ensuring associativity.

\subsection{A different rank~2 construction: foliation of $\R^4$ by
  radius-depend\-ent general-ized Connes\---Landi spheres.}

Thus far we have implicitly supposed that the transformations required
to bring $\Th(x)$ into the canonical form do not change the topology
of~$\R^4$, in regard to foliations. Star products induced by periodic
actions within the Rieffel's recipe partially escape this situation.
To illustrate this point, let us give an example of a non-constant
star product based on a different decomposition of $\R^4$, namely the
foliation of $\R^4$ by radius-dependent generalized Connes--Landi
spheres~\cite{ConnesLa}. This construction also has rank~2 and,
although lying outside the main line of this paper, we include it
because of its novelty. On ordinary space $\R^4$ define double
cylindrical coordinates $(R,\kappa,\rho,\phi)$ by
\begin{align*}
      x_1 &= R\cos\kappa\cos\phi \quad x_4 = R\sin\kappa\cos\phi\\
      x_2 &= R\cos\rho\sin\phi \quad   x_3 = R\sin\rho\sin\phi\,,
\end{align*}
where $0\leq R<+\infty,~0\leq \kappa,\rho<2\pi$ and
$0\leq\phi\leq\pi/2$. Then consider a foliation
$\R^+\!\rtimes\Sf_\omega^3$ of~$\R^4$ by spheres
of the Connes--Landi type (a continuous field of unital algebras) with
a variable noncommutativity parameter~$\omega=\omega(R)$, a smooth and
strictly positive function. Define finally the star product through
\begin{align*}
  (f\star_\omega\!h)\, (R,e^{i\kappa},e^{i\rho},\phi) & 
      = \frac{1}{ (2\pi)^2} \int\! d^2\!u\,d^2\!t\,\,e^{-iut}\\
     & ~{\scriptstyle \times}\,f(R,\,e^{i\kappa-i\omega(R)\,u_2/2},\,
                             e^{i\rho+i\omega(R)\,u_1/2},\,\phi)\,\,
        h(R,\,e^{i\kappa+it_1},\,e^{i\rho+it_2},\,\phi),
\end{align*}
where $u=(u_1,u_2)$ and $t=(t_1,t_2)$. If functions on $\R^4$ are
isotypically decomposed,
\begin{equation*}
    f=\sum_{\mathrm{\bf r}\in\Z_2}\,e^{-i(r_1\kappa+r_2\rho)}\,
      f_{\mathrm{\bf r}}(R,\phi)
\end{equation*}
the star product takes the form
\begin{equation*}
    (f \star_\omega\! h)\,(R,\kappa,\rho,\phi) 
         = \sum_{\mathrm{\bf r},\mathrm{\bf s}\in\Z_2}
         e^{-i(r_1+s_1)\kappa}\,e^{-i(r_2+s_2)\rho}\,
         e^{i\omega(R)\,\mathrm{\bf s}S\mathrm{\bf r}}\,
         f_{\mathrm{\bf r}}(R,\phi)\,h_{\mathrm{\bf s}}(R,\phi)\,.
\end{equation*}
The same analytical properties as for the $\R^2\rtimes\R^2_\theta$
foliation discussed in Subsection~2.2 and Appendix~A are easily seen
to also hold for this product.

\section{Traveling noncommutative lumps.}

The non-constant product $\star_\th$ defined in Section 2 makes
possible to extend the construction by Gopakumar, Minwalla and
Strominger~\cite{Gopakumar} of noncommutative solitons to include
propagating lumps. Let us recall in this regard that
ref.~\cite{Gopakumar} considers the extrema problem for the functional
\begin{equation*}
   V[\phi] = \int d^{2N}\!x\,\bigg(\,\frac{1}{2}~m^2 \phi^2
    + \sum_{s=3}^r \frac{b_s}{s}~ \phi_\star^s\,\bigg)\,,
\end{equation*}
where $\phi_\star^s$ is a short-hand for the noncommutative $s$-th
power
\begin{equation*}
    \phi_\star^s = \phi\star \phi\star\dots \star \phi
\end{equation*}
constructed with the constant noncommutativity matrix
\begin{equation*}
       \Th = \begin{pmatrix}  \th_1S & & & \\
                              & \th_2S & & \\
                              & & \ddots & & \\
                              & & & \th_N S
              \end{pmatrix}\,.
\end{equation*}
There it is shown that the  corresponding field equation
\begin{equation}
    m^2\phi + \sum_{s=3}^r b_s\, \phi_\star^{s-1} = 0
\label{field-equation}
\end{equation}
has solutions of the form
\begin{equation}
\phi_\Th((x_1,x_2),\ldots,(x_{N-1},x_N)) = \prod_{i=1}^N ~ 
   \sum_{k_i=0}^\infty~ a_{k_i}~f_{\th_i,k_i}(x_i,x_{i+1})\,,
\label{solution}
\end{equation}
In this expression the coefficients $a_{k_i}$ take values in the set
of solutions of the equation
\begin{equation}
   m^2 z + \sum_{s=3}^r b_s\, z^{s-1} = 0\,,
\label{algebraic}
\end{equation}
and the $f_{\th_i,k_i}$ are given by
\begin{equation}
   f_{\th_i,k_i}(x_i,x_{i+1}) = (-1)^{k_i}~2~
   e^{-\rho_i^2/\th_i}~ L_{k_i}\bigg(\frac{2\rho_i^2}{\th_i}\bigg)\,,
\label{Wigner}
\end{equation}
with $\rho_i$ defined in terms of the Cartesian coordinates
$(x_1,\ldots,x_{2N})$ as
\begin{equation*}
   \rho_{\frac{i+1}{2}}^2 = x_i^2 + x_{i+1}^2\,,\qquad
   i=1,3,\ldots,2N-1\,,
\end{equation*}
and $L_k(x)$ denoting the ordinary Laguerre polynomial of order $k$.
These solutions can be written in many different fashions, for it is
enough $\{f_{\th_i,k_i}\}$ to be a family of real orthogonal
projectors of the two-dimensional Moyal algebra with parameter
$\th_i$. The choice in eq.~\eqref{Wigner} corresponds to taking for
such projectors the Wigner eigenfunctions of the harmonic
oscillator~\cite{Groenewold,BMoyal,PhobosandDeimos}. It is also
possible to construct radial solutions for $\th_1=\cdots=\th_N=\th$ of
the form
\begin{equation}
   \phi_\th(\rho) = \sum_M a_M\,(-1)^{M}~ 
      2^N~e^{-\rho^2/\th}~
      L^{N-1}_M\bigg(\frac{2\rho^2}{\th}\bigg), \qquad M=0,1,2,\ldots,
\label{solution-radial}
\end{equation}
where $\rho$ is the radial coordinate in $\R^{2N}$ and the coefficient
$a_M$ takes values in the set of solutions of eq.~\eqref{algebraic}.

We now consider $(2N+2)$-dimensional Minkowski space-time, with
coordinates $(t,x,x_1,\ldots,x_{2N})$. From our analysis in Section 2 it
is clear that $\phi_\Th$ in eq.~\eqref{solution} is still a solution
in such a space-time of the field equation~\eqref{field-equation} when
the noncommutative parameters~$\th_i$ become functions of~$t$ and~$x$.
Furthermore, if each $\th_i$ only depends on either the sum $t+x$ or
the difference $t-x$, the field configuration $\phi_\Th$ solves the
equation
\begin{equation}
  (\pa_t^2 - \pa_x^2) \phi + m^2\phi 
    + \sum_{s=3}^r b_s\, \phi_\star^{s-1} = 0\,.
\label{improved}
\end{equation}
Such solutions describe lumps traveling in the negative or positive
directions of~$x$ with velocity~1. It is also obvious that
$\phi_\th(\rho)$ in~\eqref{solution-radial} describes a spherical
traveling lump for~$\th$ a function of $t\pm x$. However, the
solutions to eq.~\eqref{improved} discussed here are not solutions of
the field equation that results from adding to the functional
$V[\phi]$ the fully-fledged kinetic term~$-\half\,(\pa\phi)^2$.

\section{Field theory for noncommutative $\la\phi^4$ 
with variable $\th$. The effective action at one loop: generalities.}

Let us next consider quantization of a classical field theory for the
non-constant $\th(\bx)$ star product constructed in Section 2. Our
starting point for this is a real field $\phi$ with classical action
(free energy) in Euclidean space given by
\begin{equation*}
      S[\phi] = \int d^4\!x\,\bigg[\,
        \frac{1}{2} \phi\, \big(\,\tri + m^2\,\big)\, \phi
      + \frac{\lambda}{4!}~\phi\star\phi\star\phi\star\phi\,\bigg]\,.
\end{equation*}
Here $\tri$ is the Laplacian, $m^2$ is the mass squared of the field,
$\la$ is the coupling constant and we have written $\star$ instead of
$\star_\th$ to lighten the notation.  In our conventions the Laplacian
is positive: $\tri=-\pa^\m\pa_\m$.  This action defines the classical
theory.  We assume the quantum theory to be defined by its generating
path integral~\cite{IZ}
\begin{equation}
    Z[j] = {\cal N}\int [d\phi]~
        {\rm exp}\bigg(\!-S[\phi] + \int\! d^4\!x~j\phi\bigg)\,,
\label{generating}
\end{equation}
where $j(x)$ is the source with respect to which functional
differentiation of $Z[j]$ generates Green functions, and ${\cal N}$ is
a normalization factor such that $Z[0]=1$.  The path integral for
$Z[j]$ is to be understood as the formal perturbative series that
results from expanding about the classical solutions $\vf$ reducing to
the trivial one when $j=0$ of the equation
\begin{equation*}
    {\frac{\dl S[\phi]}{\dl\phi(x)}\bigg|}_{\phi=\vf} = j(x)\,.
\end{equation*}
We may suppose $\vf$ is of Schwartz class. Now, shifting in the
vicinity of a solution $\vf$ the integration variable
$\phi\to\vf+\phi$ and substituting in $Z[j]$, we have
\begin{equation*}
     Z[j] = {\cal N}\, e^{-I_0[\vf,j]} \,\int[d\phi]~
            e^{\,-\,\big(I_2[\vf,\phi]\,+\,O[\phi^3]\big)}\,,
\end{equation*}
where
\begin{eqnarray}
    &{\ds I_0[\vf,j] = S[\vf] - \int\! d^4\!x~j\vf}, &
    \cr\nonumber
    &{\ds I_2[\vf,\phi] = \int\! d^4\!x\,
      \biggl[\frac{1}{2}\,\phi\,(\tri +  m^2)\, \phi 
       + \frac{\la}{12}~\big(2\,\vf\star\vf\star\phi\star\phi 
       + \vf\star\phi\star\vf\star\phi \big)}\biggr]. &\nonumber
\end{eqnarray}
At first order in perturbation theory only the quadratic part~$I_2$ in
the exponential contributes to the path integral.

In what follows we will use the notation
\begin{equation*}
\begin{array}{c}
     H_0 = \tri + m^2, \quad H=H_0 + M\\[6pt]
     {\ds M = \frac{\la}{6} \big( L_{\vf\star\vf} + R_{\vf\star\vf}
                           + R_{\vf}L_{\vf} \big)\,, }
\end{array}
\end{equation*}
with $L_f$ and $R_f$ denoting left and right star multiplication by
$f$.  The operators $L_f$ and $R_f$ commute with each other. 
Furthermore, because $\vf$ is real, and thanks to the trace property,
$L_\vf$ and $R_\vf$ are selfadjoint.  Hence
\begin{equation*}
   L_{\vf\star\vf} + R_{\vf\star\vf} + R_\vf L_\vf
      = \half\, (L_\vf+R_\vf)^*\,(L_\vf+R_\vf)
      + \half\,( L_\vf^*L_\vf + R_\vf^*R_\vf)\,,
\end{equation*}
which shows strict positivity of~$H$. Integrating then over $[d\phi]$
we obtain for the one-loop correction to the effective
action\footnote{ Recalling from Section~3 that $L_\vf$ and $R_\vf$ for
$\vf$ of Schwartz class are bounded, so that $M$ is bounded, one can
generalize the arguments in ref.~\cite{Gayral} to prove a
priori that the difference $e^{-tH}-e^{-tH_0}$ is trace-class for all
$t>0$.}
\begin{equation}
   \Ga_1[\vf] = -\half \int_0^\infty\frac{dt}{t}\;
        \Tr\left(e^{-tH}-e^{-tH_0}\right)\,,
\label{effective}
\end{equation}
where the exponential $e^{-tH_0}$ comes from the normalization factor
${\cal N}$. We want to study the UV~divergences of $\Ga_1[\vf]$ and
see whether they can be subtracted by local counterterms. It is
well known that UV~singularities arise from the small $t$ expansion of
the integrand in eq.~(\ref{effective}). To regulate this expansion we
therefore introduce a zeta function-like regularization and consider
instead
\begin{equation}
    \Ga_1[\vf] = - \half\int_0^\infty
      \frac{\m^{2\ep}dt}{t^{1-\ep}}\,
         \Tr\left(e^{-tH}-e^{-tH_0}\right)\,.
\label{reg-effective}
\end{equation}
We must find a suitable method to compute the singularities that occur
at $\ep\to 0$ when expanding in powers of $t$ the integrand in this
expression. Because of the presence of the `background field'
$\th(\bx)$ we work in configuration space.

In position space two methods have been proposed to compute the UV
divergences for constant $\th$.  The first one, employed in
refs.~\cite{Gayral,GI}, computes $e^{-tH}$ using
Campbell--Baker--Hausdorff (CBH) or Zassenhaus-type formulae.
However, this does not work well for non-constant $\th(\bx)$. Instead
we use the method that in the physics literature goes under the name
of covariant perturbation theory~\cite{covariant}, whereas to
mathematicians is known as the Duhamel or ``expansional''
series~\cite{Araki}. For the sake of completeness,we sketch a two-line
derivation for it.  Since the standard power series for the
exponential $e^{-t(H_0+M)}$ is of little use as~$H_0$ and~$M$ do not
commute, it is better to think of~$e^{-t(H_0+M)}$ as the solution of
the differential equation
\begin{equation*}
\frac{dY(t)}{dt} = -Y(t)\,(H_0+M),
\end{equation*}
with initial condition $Y(0)=1$. Writing $\,Y(t)= X(t)\,e^{-tH_0}$, we
obtain for~$X$ the new differential equation
\begin{equation*}
\frac{dX(t)}{dt} = -X(t)M(t),
\end{equation*}
with $M(t)=e^{-tH_0}Me^{tH_0}$. Its iterative solution reads
\begin{equation*}
   e^{-tH} = \sum_{n=0}^\infty K_n(t)\,,
\end{equation*}
where $K_0(t)=e^{-tH_0}$ and 
\begin{equation*}
   K_n(t) = (-)^n\int_0^t\,d\sigma_n\int_0^{\sigma_n}d\sigma_{n-1}
     \cdots \int_0^{\sigma_2}d\sigma_1\,e^{-\sigma_1H_0}M
     e^{-(\sigma_2-\sigma_1)H_0} \cdots Me^{-(t-\sigma_n)H_0}
\end{equation*}
for $n\ge1$. Simple changes of variables~\cite{covariant} yield for
the traces of the operators $K_n(t)$
\begin{equation}
    \Tr K_n(t) =  \frac{(-t)^n}{n} \int_0^\infty d\a_1\ldots d\a_n\,
       \dl\Bigl(1-\sum\a_i\Bigr) \Tr \Big(Me^{-t\a_1H_0}Me^{-t\a_2H_0}
         \cdots Me^{-t\a_nH_0}\Big)\,,
\label{eq:traceKn}
\end{equation}
so that only $n-1$ integrals need be performed to compute $\Tr
K_n(t)$. Furthermore, since $M$ is positive and bounded from above, it
follows that
\begin{equation}
     \Tr K_n(t) \sim t^{n-2}\sepword{as} t\downarrow 0\,.
\label{estimate}
\end{equation}
Hence, only 
\begin{equation}
    \Tr K_1 = -t\Tr Me^{-tH_0}
\label{eq:K1}
\end{equation}
and
\begin{equation}
    \Tr K_2 = \half~t^2\int_0^1 d\sigma\, 
         \Tr\big[M\,e^{-t\sigma H_0}\,M\,e^{-t(1-\sigma)H_0}\big]
\label{eq:K2}
\end{equation}
will contribute to the divergent part of the regularized effective
action as $\ep\to 0$. These two contributions correspond precisely to
the two and four-point 1PI Green functions.

Because the procedure seems not as well known as perhaps it should be,
we summarize the advantages of the covariant perturbation method:
\begin{itemize}
\item CBH-like series only converge for small~$t$, whereas the
convergence properties of the Duhamel perturbation expansion are so
good that it not only adds up when $M$ is a bounded perturbation of an
unbounded~$A$ (the case here), but also for unbounded perturbations
under appropriate circumstances.
\item Small $t$ expansions by definition contain no information about
$t\!\uparrow\!\infty$.  In fact, to study $t\uparrow\infty$, one should
regard Schwinger--DeWitt formulae as asymptotic expansions in the
inverse mass squared, as it is the $e^{-tm^2}$ dampening factor that
protects the integral~(\ref{reg-effective}) from infrared problems.
The question has been tackled in the context of covariant perturbation
theory~\cite{covariant} for ordinary quantum field theory and it has
been proved that in four dimensions, irrespectively of mass, each term
$\Tr K_n$ (for $n\ge1$) of the expansion goes as $1/t$ like
$t\uparrow\infty$.  Then the effective action always exists. To our
knowledge, this question is open for NCFT.
\item The Duhamel series has a cohomological meaning. Each of its
terms is a cocycle in the entire cyclic cohomology of the Connes
bicomplex, namely the Jaffe--Lesniewski--Osterwalder
cocycle~\cite{ConnesBook, Polaris}.
\item Because the Duhamel expansion is essentially an expansion in powers
of~$M$, and $M$ is of order two in~$\vf$, each term gives directly a
$2n$-point 1PI function. From this point of view, the $\a_i$-parameters
are essentially Schwinger parameters for the propagators.
\end{itemize}

There is a price to pay for all these benefits, and that is
non-locality, making it more difficult to extract the singularities at
$\ep\to 0$. We are bound to meet here two different types of
non-locality. The first one concerns the non-locality of the Duhamel
development itself. This comes from the entanglement of the heat
kernel semigroups with the effective potentials. In the NCFT case with
constant noncommutativity, it is well known that there are non-local
contributions from both the planar and non-planar sectors of the
theory. We expect non-localities of this type to also occur here,
may be worsened now by variability of~$\th(\bx)$.

\section{The two-point sector of the effective action.}

We have already mentioned in the last section that only $K_1(t)$ and
$K_2(t)$ generate singularities at $\ep\to 0$, so that for the
regularized effective action~(\ref{reg-effective}) we can write
\begin{equation*}
   \Ga_1[\vf] = \Ga^{(2)}_1[\vf] +  \Ga^{(4)}_1[\vf]
                  + O(\ep) \,,
\end{equation*}
where
\begin{equation*}
   \Ga^{(2n)}_1[\vf] = -\,\half \int_0^\infty
       \frac{\m^{2\ep}dt}{t^{1-\ep}}~\Tr K_n(t) \qquad n=1,2.
\end{equation*}
We compute $\Ga^{\ep,(2)}_1[\vf]$ in this section and postpone
$\Ga^{\ep,(4)}_1[\vf]$ to Section 6. Using for $\Tr K_1(t)$ the
expression in~\eqref{eq:K1}, we have
\begin{equation}
    \Ga^{(2)}_1[\vf] = \frac{\la\,\m^{2\ep}}{12} \int_0^\infty
    dt\,t^\ep e^{-tm^2}\,\Tr\Big[(L_{\vf\star\vf} + R_{\vf\star\vf}
                      + R_\vf L_\vf)\, e^{-t\tri}\Big]\,.
\label{action-2}
\end{equation}

We first calculate the traces of the three terms in the integrand
and then integrate over $dt$. The $L_{\vf\star\vf}$ and
$R_{\vf\star\vf}$ terms in eq.~(\ref{action-2}) yield the same
contribution to the trace, since for any function $f$ one has
\begin{equation}
    \Tr\big( R_f\,e^{-t\tri}\big)  = \Tr \big(L_f\,e^{-t\tri}\big)
         = \Tr \big(M_f\,e^{-t\tri}\big)
         =  \frac{1}{(4\pi t)^2}\int\! d^4\!x ~ f(\bx,\tx) \,,
\label{L-R-M}
\end{equation}
where $M_f$ denotes the operator of ordinary multiplication by~$f$.
To prove this, we note that from the definition of star product in
eq.~(\ref{rank2}) it follows, in obvious notation, that
\begin{eqnarray}
    \langle x|L_f |y\rangle  &\igual&  \frac{1}{(2\pi)^2} \int
        d^2\!\tu\,d^2\!\tv~ e^{-i\tu\tv}~
        f\big(\bx,\,\tx-\half\,\th(\bx)\,S\tu\big)~
        \dl^{(2)}(\bx-\by)~\dl^{(2)}(\tx+\tv-\ty) \label{xLy}\\
    \langle x|R_f |y\rangle &\igual& \frac{1}{(2\pi)^2} \int
        d^2\!\tu\,d^2\!\tv~ e^{-i\tu\tv}~ f\big(\bx,\,\tx+\tv)~
        \dl^{(2)}(\bx-\by)~
        \dl^{(2)}\big(\tx-\half\th(\bx)S\tu-\ty \big) \,. \label{xRy}
\end{eqnarray}
Taking e.g. $L_f$ and noting
\begin{equation*}
   \langle x|e^{-t\tri}|y\rangle = \frac{1}{(4\pi t)^2}~
                 e^{-|x-y|^2/4t}\,,
\end{equation*}
we write
\begin{eqnarray*}
    \Tr\big( L_f\,e^{-t\tri}\big)
      &\igual & \int\! d^4\!x\, d^4\!y~ \langle x|L_f |y\rangle~
        \langle y|e^{-t\tri}|x\rangle  \\
      &\igual & \frac{1}{(2\pi)^2} \int\! d^4\!x \int\! d^2\!\tu\,
       d^2\!\tv~e^{-i\tu\tv}~f\big(\bx,\,\tx-\half\,\th(\bx)\,S\tu\big)~
      \frac{e^{-\tv^2/4t}}{(4\pi t)^2}\,.
\end{eqnarray*}
Making now the change $\,\tx\to \tx+\half\,\th(\bx)\,S\tu\,$ and
integrating over $\tu$ and $\tv$, we obtain
\begin{equation*}
    \Tr\big( L_f\,e^{-t\tri}\big)
       =  \frac{1}{(4\pi t)^2}\int\! d^4\!x ~ f(\bx,\tx) \,.
\end{equation*}
The proof for $R_f$ goes along the same lines. Next we move on to the
the third term in eq.~(\ref{action-2}). Using similar arguments, it is
straightforward to show that
\begin{equation}
    \Tr\big(R_f\,L_f\,e^{-t\tri}\big) = \frac{1}{(2\pi)^2}
       \int \frac{d^4\!x}{\th^2(\bx)} \int\! d^2\!\tu~
         f(\bx,\tx)~f(\bx,\tx+\tu)~
           \frac{e^{-t\tu^2/\th^2(\bx)}}{4\pi t}\,.
\label{RL}
\end{equation}
Substituting the results~(\ref{L-R-M}) and~(\ref{RL}) in
eq.~(\ref{action-2}) and integrating over~$t$ we have
\begin{eqnarray}
    \Ga^{(2)}_1[\vf]  &\igual &\frac{\la m^2}{96\pi^2}\,
       \left(\frac{\m^2}{m^2}\right)^{\!\ep} \,\Ga(-1+\ep)
       \int\! d^4\!x~\vf^2(x)
       \nonumber\\
	  & \mas &
	  \frac{\la}{192\pi^3}\,\left(\frac{\m^2}{m^2}\right)^{\!\ep}
	  \,\Ga(\ep)\int\!\frac{d^4\!x}{\th^2(\bx)}\int\! d^2\!\tu~
	  \vf(\bx,\tx)~\vf(\bx,\,\tx+\tu)~{\left[1
	  + \frac{\tu^2}{m^2\th^2(\bx)}\right]}^{-\ep}\,.
       \label{action-2-result}
\end{eqnarray}
The first term gives a local contribution.  Hence the divergence that
occurs in it when $\ep\to 0$ can be subtracted by a local counterterm. 
By contrast, the second term in~(\ref{action-2-result}) is non-local
and, as a result, the singularity that it develops as $\ep\to 0$ can
not be subtracted by a local counterterm, thus spoiling
renormalizability.  Although this is not the by now traditional UV/IR
mixing of NCFT, it reveals a connection between the UV and IR sectors,
since the UV singularity at~$\ep\to 0$ in~$\Ga(\ep)$ gets mixed with
the long distance correlation contained in the integral over $\tu$. 
Note also that in eq.  \eqref{action-2-result} we observe the
traditional splitting of $\th$-constant NCFT in a planar part (first
term) with the same dependence on $\th(\bx)$ as in the classical
action, and a non-planar part (second term) with a more complicated
dependence.

Except for the fact that now $\th(\bx)$ depends on $\bx$, this
non-locality of UV~divergences also occurs for $\th$-constant
NCFT. Just take $\th(\bx)=\th$ constant and the result of
ref.~\cite{GGBRR} is recovered. There it is also explained how
the corresponding result in momentum space ---see e.g.~\cite{FRR}---
conceals the singularity of the $\ep\to 0$ limit. For constant~$\th$
one may think of avoiding the problem by considering noncommutativity
matrices of rank~4~\cite{Bahnsetalt, Fujikawa} or by supersymmetrizing
the theory~\cite{FRRsusy, Matusis}. In the non-constant $\Th(x)$ case,
as seen in Section 3, associativity excludes rank~4 noncommutativity
matrices. Supersymmetry remains unexplored.

When several vertices come into play, in the trace computations one
has to deal with (sums and differences of) the noncommutativity
function $\th(\bx)$ evaluated at different points. This will mix
further the UV and IR sectors of the theory. To better understand this
mixing, and because we do not yet want to exclude non-local UV
renormalizations in a theory with a non-local $x$-dependent background
$\th(\bx)$, which most likely is to be understood as an effective
theory, we forge ahead, and undertake the calculation of the
four-point part of the effective action. This will also allow us to
discuss methods to deal with the apparent non-locality of the Duhamel
expansion.

\section{The four-point sector of the effective action.}

In what follows we study $\Ga_1^{(4)}$.  Using~\eqref{eq:K2} we have
\begin{equation*}
    \Ga_1^{(4)} = -\frac{\m^{2\ep}}{4}\int_0^\infty\!dt\,t^{1+\ep}
       \int_0^1\!d\sigma\, \Tr\big[M\,e^{-t\sigma H_0}\,
          M\,e^{-t(1-\sigma)H_0}\big]\,.
\end{equation*}
Recalling the expression of $M$ in terms of left and right
multiplication operators, using the invariance of the integral under
$\sigma\to1-\sigma$, cyclicity of the trace and the property
\begin{equation*}
     \langle x|R_f|y\rangle =  \langle x|L_{f^*}|y\rangle^*
         = \langle y|L_f|x\rangle\,,
\end{equation*}
we obtain for $\Ga_1^{(4)}$ the sum of four terms
\begin{equation}
    \Ga_{X_I}^{(4)} =
    -\frac{\,\la^2\m^{2\ep}\,}{144}\int_0^\infty\!dt\,
        t^{1+\ep} \int_0^1\!d\sigma\, X_I(t,\sigma) \qquad I=1,\ldots,4
\label{1234}
\end{equation}
with
\begin{align*}
      X_1 &= 2\Tr\big[ L_{\vf\star\vf}\,e^{-t\sigma H_0}\,
             L_{\vf\star\vf}\,e^{-t(1-\sigma)H_0}\big] \\
      X_2 &= 2\Tr\big[ L_{\vf\star\vf}\,e^{-t\sigma H_0}\,
             R_{\vf\star\vf}\,e^{-t(1-\sigma)\,H_0}\big] \\
      X_3 &= 4\Tr\big[ L_{\vf\star\vf}\,e^{-t\sigma H_0}\,
             R_{\vf}\,L_{\vf}\,e^{-t(1-\sigma)H_0}\big] \\
      X_4 &= \Tr\big[ R_{\vf}L_{\vf}\,e^{-t\sigma H_0}\,
             R_{\vf}\,L_{\vf}\,e^{-t(1-\sigma)H_0}\big] \,.
\end{align*}
We shall see below that $\Ga^{(4)}_{X_1}$ becomes singular at
$\ep\to 0$, whereas $\Ga^{(4)}_{X_I}$ is finite for $I=2,3,4$.

\subsection{UV divergences.}

Let us first consider $\Ga^{(4)}_{X_1}$. Using the expressions for
the matrix elements of $L_f$ and $R_f$ in eqs.~(\ref{xLy})
and~(\ref{xRy}), it is straightforward to see that~$X_1$ can be recast
as
\begin{align*}
    X_1 &= \frac{1}{(2\pi)^4} \int d^4\!x\,d^4\!z
       \int\!d^2\tu\,d^2\tv\,d^2\tu'\,d^2\tv'\,e^{-i(\tu\tv+\tu'\tv)}~
       \big(\vf\star\vf\big) \big(x-\half\,\th(\bx)\,S\tu\big)\\
    &~{\sc \times}~
       \big(\vf\star\vf\big) \big(z-\half\,\th(\bz)\,S\tu'\big)~
       \langle x+\tv|e^{-t\sigma H_0}|z\rangle~
       \langle z+\tv'|e^{-t(1-\sigma)H_0}|x\rangle\,.
\end{align*}
We now make the changes
\begin{equation*}
     \tx \to \tx+\half\,\th(\bx)\,S\tu \qquad
     \tv \to \tv-\half\,\th(\bx)\,S\tu \qquad
     \tv' \to \tv'+\half\,\th(\bx)\,S\tu \,,
\end{equation*}
integrate over $\tu$ and $\tv$ and in the result perform 
\begin{equation*}
     \tz \to \tz +\half\,\th(\bz)\,S\tu' \qquad
     \tv' \to \tv'-\half\,\th(\bz)\,S\tu' \,.
\end{equation*}
Proceeding in this way we obtain
\begin{align}
     X_1 & = \frac{1}{(2\pi)^2} \int d^4\!x\,d^4\!z\int\!d^2\tu\,d^2\tv~
       e^{-i\tu\tv}~ (\vf\star\vf)(x)~(\vf\star\vf)(z) \nonumber \\
     &~{\sc \times}~
       \big\langle x+\half\,\big[\th(\bx)-\th(\bz)\big]S\tu\,
           |e^{-t\sigma H_0}|z\big\rangle~
       \langle z+\tv|e^{-t(1-\sigma)H_0}|x\rangle\,.
\label{X1}
\end{align}
Doing now the Gaussian integrations over $\tu$ and $\tv$ and
substituting in the expression for~$\Ga^{(4)}_{X_1}$, we arrive at
\begin{equation}
    \Ga_{X_1}^{(4)} = -\frac{\,\la^2\m^{2\ep}\,}{72\,(2\pi)^4}
      \int d^4\!x\, d^4\!z \int_0^\infty\!dt~ t^{-1+\ep}~e^{-tm^2}
      \int_0^1\!d\sigma~ (\vf\star\vf)(x)~ (\vf\star\vf)(z)~
      K(x,z;t,\sigma)
\label{th-th-int}
\end{equation}
where the kernel $K(x,z;\,t,\sigma)$ is given by
\begin{equation}
\begin{array}{rl}
  K(x,z;\,t,\sigma) = &\!\! {\ds \frac{1}{4t\sigma(1-\sigma)}~ 
    {\rm exp}\, \bigg[\! - \frac{(\bx-\bz)^2}{4t\sigma(1-\sigma)}\, 
                \bigg]} \\[9pt]
    {\sc \times} &\!\! {\ds \frac{1}
                 {\,4t\sigma(1-\sigma) + \th^2_-(\bx,\bz)/t\,}~    
    {\rm exp}\,\bigg[\! - \frac{(\tx-\tz)^2}{4t\sigma(1-\sigma)
                       + \th^2_-(\bx,\bz)/t\,}\, \bigg]\,,}
\end{array}
\label{kernel}
\end{equation}
and by definition
\begin{equation*}
    \th_\pm(\bx,\bz) = \half\,\big[\, \th(\bx)\pm\th(\bz)\, \big]\,.
\end{equation*}
To analyze this expression we consider first the case in which
$\th_-=0$. Eq.~\eqref{th-th-int} then takes the form
\begin{equation}
    \Ga_{X_1,\,\th_-=0}^{(4)} =
        -\frac{\,\la^2\m^{2\ep}}{72\,(4\pi)^2}
        \int d^4\!x\, d^4\!z~\vf\star\vf(x) \int_0^\infty\!dt~
        \frac{e^{-tm^2}}{t^{1-\ep}} \int_0^1\!d\sigma~
        \frac{e^{-|x-z|^2/4t\sigma(1-\sigma)}}
             {[4\pi t\sigma(1-\sigma)]^2}~
        \vf\star\vf(z)\,.
\label{surprising}
\end{equation}
This integral can be written as
\begin{equation*}
   \Ga_{X_1,\,\th_-=0}^{(4)} =
      -\frac{\la^2\m^{2\ep}}{\,72\,(4\pi)^2}
      \int d^4\!x \int_0^\infty\!dt~
      \frac{e^{-tm^2}}{t^{1-\ep}} \int_0^1\!d\sigma~
      (\vf\star\vf)(x)~ \big[ e^{-t\sigma(1-\sigma)\tri}\,
      (\vf\star\vf)\big]\,(x)\,.
\end{equation*}
Integrating over~$t$, expanding in powers of $\ep$ and integrating
over~$\sigma$, we obtain
\begin{equation*}
   \Ga_{X_1,\,\th_-=0}^{(4)}
   = -\frac{\la^2}{\,72\,(4\pi)^2} \int d^4\!x~
        (\vf\star\vf)\,(x) \bigg[ \frac{1}{\ep} -\ga 
            - \ln\Big(\frac{m^2+\tri/6}{\m^2}\Big) +O(\ep)\bigg]\,
        (\vf\star\vf)\,(x)\,.
\end{equation*}
This gives a local UV~divergence, which in our regularization scheme
is characterized by a pole in~$\ep$. It is worth mentioning the
non-local aspect of~\eqref{th-th-int}. Note that even if $\th=0$,
which corresponds to the ordinary commutative $\la\phi^4$ model,
formula~\eqref{surprising} looks non-local. The reason for that is the
intrinsic non-locality of the Duhamel method. This is the apparent
non-locality to which we referred at the end of Section 3. 

If $\th_-\ne0$, we divide the domain of integration for $t$ into two
parts: from $0$ to $\a$ and from $\a$ to $\infty$, with
$\a>0$. The contribution from $\a$ to $\infty$ is finite for
$\ep\to 0$. That from $0$ to $\a$ becomes singular at
$\ep\to 0$. To find the singularity we recall the distributional
identity
\begin{equation*}
    \lim_{\a\downarrow0}~ \frac{1}{(\pi\a)^{k/2}}~ 
       e^{-x^2/\a} = \dl^{k}(x)  \sepword{for} x\in\R^{k},
\end{equation*}
where $\dl^{k}$ denotes the $k$-dimensional delta function. This gives
for the kernel $K(x,z;\,t,\sigma)$ the following behavior for small
$t$:
\begin{align*}
   \lim_{t\downarrow0}\,K(x,z;\,t,\sigma) & =
       \lim_{t\downarrow0}\, \frac{\dl^2(\bx-\bz)}
           {\,4t\sigma(1-\sigma) + \th^2_-(\bx,\bz)/t\,}~    
    {\rm exp}\,\bigg[\! - \frac{(\tx-\tz)^2}{4t\sigma(1-\sigma)
                       + \th^2_-(\bx,\bz)/t\,}\, \bigg] \\
    & =\dl^4(x-z)\,.
\end{align*}
Taking into consideration the factor $t^{-1+\ep}e^{-tm^2}$, we obtain
for the singular part of $\Ga^{(4)}_{X_1}$, after integration
over $t$ and $\sigma$,
\begin{equation}
    \Ga^{(4)}_{X_1,\,{\rm sing}} = -\frac{\la^2}{\,72\,(4\pi)^2\,\ep}
        \int d^4\!x~\big[(\vf\star\vf)\,(\vf\star\vf)\big]\,(x)\,.
\label{divergence}
\end{equation}
We thus recover the same singularity (pole part) as for the $\th_-=0$
case, which can be subtracted by adding a local counterterm to the
classical action. Note in this regard that $K(x,z;\,t,\sigma)$ has the
same behavior for small $t$ as that of the kernel
\begin{equation*}
   \frac{e^{-|x-z|^2/4t \sigma(1-\sigma)}}
        {[4\pi t\sigma(1-\sigma)]^2}
\end{equation*}
corresponding to $\th_-=0$.  Let us, nevertheless, remark that in NCFT
at constant $\th$ the contribution of~$X_1$ is purely planar, whereas
for non-constant $\th$ it is not. In fact, in eq.~\eqref{th-th-int}
the finite planar and non-planar contributions to the effective action
thoroughly mix. With this we complete the analysis of
$\Ga^{(4)}_{X_1}$ in eq.~(\ref{1234}). Let us now move to that of
$\Ga^{(4)}_{X_I}$ for $I=2,3,4$.

\subsection{Finite contributions.}

Using the expressions for the matrix elements of $L_f$ and $R_f$, and
proceeding similarly as for $X_1$, it is not difficult to obtain for
$X_2$
\begin{align}
     X_2 & = \frac{1}{(2\pi)^2} \int d^4\!x\,d^4\!z\int\!d^2\tu\,d^2\tv~
       e^{-i\tu\tv}~ (\vf\star\vf)(x)~(\vf\star\vf)(z) \nonumber \\
     &~{\sc \times}~
       \big\langle x+\th_+(\bx,\bz)S\tu\,
           |e^{-t\sigma H_0}|z\big\rangle~
       \langle z+\tv|e^{-t(1-\sigma)H_0}|x\rangle\,.
\label{cut-off}
\end{align}
This is the same expression as for $X_1$ in eq.~(\ref{X1}), with the
only difference that instead of the half-difference $\th_-(\bx,\bz)$
we have the half-sum $\th_+(\bx,\bz)$. Slightly more complicated
expressions involving $\th_+(\bx,\bz)$ can be obtained for $X_3$ and
$X_4$. Since $\th(\bx)$ is strictly positive, the half-sum will also
be strictly positive and will act in eq.~(\ref{cut-off}) as a
cut-off. This hints that the contribution from $X_2$ to the effective
action has no singularities at $\ep=0$. In fact in appendix C it is
shown that $X_2,\,X_3$ and $X_4$ are bounded from above by a constant
times $e^{-tm^2}/t$,
\begin{equation}
     |X_I| \leq C_I\,\frac{e^{-tm^2}}{t} \qquad I=2,3,4.
\label{bound}
\end{equation}
Then, upon integration over $t$ and $\sigma$, they give finite
contributions to the effective action, since
\begin{equation*}
    |\Ga^{(4)}_{X_I}| \leq C'_I\, \la^2\m^{2\ep}
        \int_0^\infty \! dt\,t^\ep\,e^{-tm^2} =
        C'_I\la^2\left(\frac{\m^2}{m^2}\right)^{\!\ep}
        \,\Ga(\ep+1) = C'_I\la^2 + O(\ep) .
\end{equation*}
All in all we conclude that the only divergence in the four-point part
of the effective action is due to $X_1$, has the
form~(\ref{divergence}) and, as already stated, can be subtracted by a
local counterterm, posing no problem to renormalizability.

Note that having a non-constant $\th(\bx)$ does not solve the
traditional UV/IR mixing. This is obvious from eq.~(\ref{cut-off}),
for if we make $\th$ approach zero, $\th_+$ also approaches zero and
the divergences due to short distances $x\to z$ reappear.

\section{Conclusion.}

In this paper we have considered the generalization within Rieffel's
formalism of the Moyal product in~$\R^4$ to noncommutative products
with position-dependent noncommutativity parameters, and the
formulation of quantum field theory for such generalized
noncommutative products.

In constructing such non-constant noncommutative products, the
requirement of associativity proves very restrictive and only allows a
rank~2 matrix $\Th(x)$ with entries $\Th^{ij}(x)=0$ and
$\Th^{ab}(x)=\eps^{ab}\th(\bx)$. Here $x=(\bx^1,\bx^2,\tx^1,\tx^2)$
and $\th(\bx)$ is an arbitrary positive sufficiently smooth bounded
function. Perhaps more interesting is to regard this noncommutative
product as a foliation $\R^2_{\bx}\rtimes\R^2_{\th(\bx)}$ of $\R^4$.
This point of view makes possible to generalize this construction
to~$n\geq 3$ dimensions by writing $\R^n\simeq\R^{n-2m}\rtimes
\R^{2m}$. The existence of these `non-constant' star products in
dimension $n$, where noncommutativity is constrained to $2m$
dimensions and depends on the other $n-2m$ coordinates allows to
generalize the existing noncommutative solitons in the
literature~\cite{Gopakumar} to traveling noncommutative lumps.  In
this regard, it would be interesting to investigate its usefulness in
the search for other time-varying noncommutative classical solutions.
It also suggests to consider them in connection with extra
dimensions. An interesting question concerns the interpretation of
variable noncommutativity in relation with the quantum group approach
to the Moyal product~\cite{Oeckl}.

To do quantum field theory on our noncommutative $x$-dependent
background, we have worked in position space and used the covariant
perturbation method~\cite{covariant}. We have been able to perform
explicit calculations at one loop for the $\la\phi^4$ model. Our
computations for the two and four-point 1PI functions reveal that
quantum field theory for non-constant $\th(\bx)$ essentially suffers
from the same problems as for constant $\th$. In this regard, the
arbitrariness in the choice of $\th(\bx)$ has not been of much use and
the situation is essentially not better nor worse than for
constant~$\th$.

We close by mentioning that the construction of star products with
non-constant $\Th(x)$ in Minkowski space-time is more involved, as the
orbits of $\Th$ under the action of the Lorentz group have a more
complicated structure than under orthogonal similarity. The
construction found here provides an $x$-dependent generalization of a
space-like or magnetic constant noncommutativity in Minkowski
space-time, for which the effective action can be obtained by analytic
continuation.

\subsection*{Acknowledgment}

The authors thank J. C. V\'arilly for helpful comments. VG wishes to
acknowledge the hospitality of the Department of Theoretical Physics
of Univer\-sidad Complutense de Madrid, where this work was started.
JMGB is very grateful to B. Booss-Bavnbek for making available to him
unpublished notes on the Duhamel expansion. He also thanks MEC, Spain
for support through a `Ram\'on y Cajal' contract. FRR is grateful to
MEC, Spain for financial support through grant No.~BFM2002-00950.

\section*{Appendix A: Algebra property and boundedness of left and
   right multiplication. }
\renewcommand{\theequation}{A.\arabic{equation}}

We collect in this appendix other basic properties 
of the star product $\star_\theta$ defined in Subsection 2.2. 

1. Involutivity. From  the definition given
   in eq.~(\ref{rank2}) it is obvious that
\begin{equation*}
   \bigl(f \star_\th g\bigr)^* =  g^* \star_\th f^*\,.
\end{equation*}

2. Equivariance. For arbitrary $\tilde{u}\in\R^2$
and~$\,M(x)\in SL(2,\R)$, the following can be seen to hold:
\begin{align*}
    (f\star_\theta g)\,\,(\bx,\,\tx+\tilde{u})
       & = f(\bx,\,\tx+\tilde{u}) \star_\th g(\bx,\,\tx+\tilde{u}) \\
    (f\star_\theta g)\,(\bx,\,M(x)\,\tx)
       & = f(\bx,\,M(x)\,\tx)\star_\theta g(\bx,\,M(x)\,\tx)\,.
\end{align*}

3. Algebra property. We have implicitly assumed that the
functions $f,\,g,\,\ldots$ for which the star product $\star_\theta$
is defined and the function $\theta(\bx)$ satisfy good enough
properties for the different objects that we have considered to be
well-defined. Let us now be more precise about this point. By
$\SS(\R^4)$ we denote the space of Schwartz (smooth, rapidly
decreasing together with all derivatives) functions on~$\R^4$. We
recall that the standard Fr\'echet topology of $\SS(\R^4)$ is given by
the family of semi-norms
\begin{equation*}
    p_{_{k,l}}(f)= \sup_{|\a|\leq k,\,|\b|\leq l}~
       \sup_{x\in\R^4}\left|x^\a\pa^\b f(x)\right|,
\end{equation*}
or alternatively by 
\begin{equation*}
    q_{_{k,l}}(f)=\sup_{|\a|\leq k,\,|\b|\leq l}\,\,
        \sup_{\bx\in\R^2} \left\{\int d^2\!\tx
             \left|x^\a\pa^\b f(x)\right|\right\}.
\end{equation*}
Since $\th(\bx)>0$, we can rewrite the star product as
\begin{equation*}
    (f\star_\th\! g)\,(\bx,\tx )= \frac{1}{\pi^2\,\th^2(\bx)}
    \int d^2\!\ty\,d^2\!\tz~ e^{-2i\,(\tx-\ty)S(\tx-\tz)/\th(\bx)}~
       f(\bx,\ty)\,g(\bx,\tz).
\end{equation*}
Thus, if 
\begin{equation}
    \frac{C_1}{(1+|\bx|^2)^{k_0}} \leq \theta(\bx) \leq C_2 \,,
\label{condition}
\end{equation}
with $k_0$ a positive integer and and $C_1,C_2>0$ constants, we get
\begin{eqnarray*}
   p_{_{0,0}}(f\star_\th g) & \menorigual & \frac{1}{\pi^2}\,
     \sup_{\bx\in\R^2} \left\{\frac{1}{\th(\bx)} \int d^2\!\ty
                              \left|f(\bx,\ty)\right| \right\}
     \sup_{\bx\in\R^2} \left\{\frac{1}{\th(\bx)} \int d^2\!\tz
                              \left|g(\bx,\tz)\right| \right\}\\
   &\menorigual & \frac{1}{C_1^2\,\pi^2}
     \bigg[ \sum_{j=0}^{2k_0} \binom{2k_0}{j}~q_{_{2j,0}}(f)\bigg]\,
     \bigg[ \sum_{i=0}^{2k_0} \binom{2k_0}{i}~q_{_{2i,0}}(g)\bigg]\,.
\end{eqnarray*}
For other values of $k$ and $l$, we obtain similar estimates, provided
all the derivatives of the function $\th(\bx)$ are polynomially bounded.
Indeed, noting
\begin{eqnarray*}
     \bx^i \big(f\star_\th g\big)(\bx,\tx) &\igual &
         (\bx^i f) \star_\th g(\bx,\tx)
       = f\star_\th (\bx^i g)(\bx,\tx) \\[3pt]
     \tx^a \big(f\star_\th g\big)(\bx,\tx) &\igual &
         f\star_\th (\tx^a g)(\bx,\tx)
       + \ihalf\,\th(\bx)\,\eps^{ab}\,
         \Big(\frac{\pa f}{\pa\tx^b}\star_\th g\Big) (\bx,\tx) \\
     &\igual & (\tx^a f) \star_\th g(\bx,\tx)
       - \ihalf\,\th(\bx)\,\eps^{ab}\,
         \Big( f\star_\th \frac{\pa g}{\pa\tx^b} \Big)(\bx,\tx),
\end{eqnarray*}
and using~(\ref{LeibnizOK}) and~(\ref{Leibnizviol}), we obtain by
induction
\begin{equation*}
     p_{_{k,l}}(f\star_\th g) \leq \sum_{i\in I,\,j\in J}
	 K(C_1,k_0,k'_i,k'_j)\,
        q_{_{k'_j,k_j}}(f)~q_{_{k'_i,k_i}}(g),
\end{equation*}
for some finite sets $I,J$. This shows that the star product
$\star_\th$ is separately continuous, indeed jointly continuous, since
$\SS(\R^4)$ is a Fr\'echet space.  Hence the space $\SS(\R^4)$ endowed
with the star product~$\star_\theta$ is an associative and involutive
Fr\'echet algebra with jointly continuous product.

Under the same conditions the operators of left~$L_f$ and right~$R_f$
star multiplication by $f$ are bounded on the Hilbert
space~$L^2(\R^4)$. For $L_f$ this means that $\|L_{\!f}\psi\|=
\|f\star_\th \psi\|<\infty$ for all $f$ in $\SS(\R^4)$ and all $\psi$
in~$L^2(\R^4)$. To prove it, we make in eq.~\eqref{rank2} the change
$\tu\to 2\tu/\th(\bx)$, use the identity
\begin{equation}
    e^{-2i\tu\tv/\th(\bx)}
      = \frac{1}{(1+|\tu|^2)^k}~
        \Big[1+ \fourth\,\th^2(\bx)\,\tri_{\tv}\Big]^k\,
        e^{-2i\tu\tv/\th(\bx)}
\label{oscillatory}
\end{equation}
for some integer $k>1$, and integrate by parts. For all $\psi\in
L^2(\R^4)$ we then have, after further changing $\tu\to \tx
-S^{-1}\tu$, 
\begin{equation*}
   \left|L_f \psi(\bx,\tx)\right| \le \frac{1}{\pi^2\,\th^2(\bx)} 
   \int d^2\!\tu\,d^2\!\tv~ 
   \Big| \Big[(1+\fourth\,\th^2(\bx)\,\tri_{\tu})^k f\Big] 
         (\bx,\tu) \Big|~ 
    \frac{ \big|(U_{\tv}\psi)(\bx,\tx)\big|}{(1+|\tv|^2)^k}~,
\end{equation*}
with $U_{\tv}$ the unitary operator of translation by $\tv$.  Finally,
we obtain
\begin{equation*}
   \|L_f\psi\|_2 \leq \frac{1}{\pi^2}~ \sup_{\bx\in\R^2}
     \left\{ \frac{1}{\th^2(\bx)} \int d^2\!\tu\,
           \left|\, \Big[(1+\fourth\,\th^2(\bx)\,\tri_{\tu})^k f\Big]
                     (\bx,\tu)\, \right| \right\}\, \|\psi\|_2\!
     \int\! d^2\!\tv\, \frac{1}{(1+|\tv|^2)^k}\,,
\end{equation*}
which is finite for~$k>1$ since $f\in\SS(\R^4)$ and~$\th(\bx)$
satisfies condition~(\ref{condition}).

In our arguments in Section 2 we used coordinate functions not
belonging to~$\SS(\R^4)$. That is a trivial rigour point, however, as
they do belong to its multiplier algebra. The reader is
referred in this regard to~\cite{PhobosandDeimos}, as everything here
works the same as in those references.

\section*{Appendix B: Proof of eq.~(\ref{bound}).}
\renewcommand{\theequation}{B.\arabic{equation}}

In this appendix we prove eq.~(\ref{bound}). We present a detailed
discussion of the more involved case, namely $X_4$. The proofs for
$X_2$ and $X_3$ go along the same lines and can be easily retrieved
from the one presented here.

Using the expressions for the matrix elements of $L_f$ and $R_f$ in
eqs.~(\ref{xLy}) and~(\ref{xRy}), we obtain after some simple changes
of variables and two plane wave integrations that
\begin{align*}
     X_4 & = \frac{1}{(2\pi)^4} \int\!d^4\!x\,d^4\!z
      \int\!d^2\tu\,d^2\tv\,d^2\tu'\,d^2\tv'\,e^{-i(\tu\tv+\tu'\tv')}\,
        \vf(x+\tv)~\vf\big(x-\half\theta(\bx)S\tu-\tv\big) \\
     &~{\sc \times} ~
       \vf(z+\tv')~\vf\big(z-\half\,\th(\bz)\,S\tu'-\tv'\big)~
       \langle x-\half\,\theta(\bx)\,S\tu|e^{-t\sigma H_0}|z\rangle~
       \langle z-\half\,\th(\bz)\,S\tu'|e^{-t(1-\sigma)H_0}|x\rangle\,.
\end{align*}
We now make the changes $\tu\to 2\tu/\th(\bx)$ and $\tu'\to
2\tu'/\th(\bz)$, use the identity \eqref{oscillatory} and integrate by
parts, noting that the action of $\,\Big[1+
\fourth\,\th^2(\bx)\,\tri_{\tv}\Big]^k\,$ on the product
$\,\vf(x+\tv)\,\vf(x-S\tu-\tv)\,$ generates a sum of the type
\begin{equation*}
    \sum_{I+J=0,\ldots,2k}
    \Big[P_I\big(\half\th(\bx)\tilde{\pa}\big)\,\vf\Big](x+\tv)~
    \Big[P_J(\half\th(\bx)\tilde{\pa})\,\vf\Big](x-S\tu-\tv)\,,
\end{equation*}
with $P_I(\cdot)$ and $P_J(\cdot)$ polynomials of order $I$ and $J$.
This yields
\begin{align*}
     |X_4| \leq \frac{1}{\pi^4} \sum_{I+J=0,\ldots,2k}\,& \int\!
        \frac{d^4\!x}{\th^2(\bx)}~\frac{d^4\!z}{\th^2(\bz)}
        \frac{d^2\!\tu\,d^2\!\tv}{(1+|\tu|^2)^k}~
        d^2\!\tu'\,d^2\!\tv'\, \\[-1pt]
     &{\sc \times}~
        \Big|P_I\big(\half\th(\bx)\tilde{\pa}\big)\,\vf(x+\tv)\Big|~
        \Big|P_J(\half\th(\bx)\tilde{\pa})\,\vf(x-S\tu-\tv)\Big|~
        \\[2pt]
     &{\sc \times}~
        \big|\vf(z+\tv')\big|~\big|\vf(z-S\tu'-\tv')\big|~
        \langle x-S\tu|e^{-t\sigma H_0}|z\rangle~
        \langle z-S\tu'|e^{-t(1-\sigma)H_0}|x\rangle\,,
\end{align*}
Making the change $\tv\to \tv-\tx$ and recalling that the norm
$\|\!\cdot\!\|_\infty$ of a function $f$ defined on 
$\R^n$ is $\|f\|_\infty = {\ds \sup_{x\in \R^n}}\,|f(x)|$, we have
\begin{align}
   |X_4| \leq \frac{1}{\pi^4}~  K(\th,k,\vf)
         & \int\!  d^4\!x~\frac{d^4\!z}{\th^2(\bz)}~
         \frac{d^2\!\tu}{(1+|\tu|^2)^k}~d^2\!\tu'~d^2\!\tv'~
         \big|\vf(z+\tv')\big| \nonumber \\[2pt]
     &~{\sc \times}~\big|\vf(z-S\tu'-\tv')\big|
        ~\langle x-S\tu|e^{-t\sigma H_0}|z\rangle~
        \langle z-S\tu'|e^{-t(1-\sigma)H_0}|x\rangle\,.
\label{boundpartial}
\end{align}
The constant $K(\th,k,\vf)$ is given 
\begin{equation*}
    K(\th,k,\vf) = \sum_{I+J=0,\ldots,k} ~
      \sup_{\bx\in \R^2} \left\{ \frac{1}{\th^2(\bx)}~
       \Big\|P_J(\half\th(\bx)\tilde{\pa})\,\vf \Big\|_\infty
       \int d^2\!\tv~ \Big|P_I\big(\half\th(\bx)\tilde{\pa}\big)\,
               \vf(\bx,\tv)\Big|~\right\}
\end{equation*}
and is a finite, since $\th(\bx)$ is finite and non rapidly decreasing
by condition~(\ref{condition}), and $\vf$ is of Schwartz class.

The semigroup property
\begin{equation*}
   \int d^4\!y~ \langle x|e^{-t_1H_0}|y\rangle~
                 \langle y|e^{-t_2H_0}|z\rangle
      =\langle x|e^{-(t_1+t_2)H_0}|z\rangle
\end{equation*}
now allows us to perform the integral over $x$ in
eq.~(\ref{boundpartial}), with the result $\langle z-S\tu'| e^{-tH_0}|
z+S\tu\rangle$. Using this and making the changes $\tz\to \tz-\tv'\,$
and $\,\tv'\to \tv'-\half \tu'\,$ we recast eq.~(\ref{boundpartial})
as
\begin{equation*}
   |X_4| \leq  \frac{1}{4\pi^4}~ K(\th,k,\vf)
        \int\!\frac{d^4\!z}{\th^2(\bz)}~
        \frac{d^2\!\tu}{(1+|\tu|^2)^k}~d^2\!\tu'~d^2\!\tv'~
         \big|\vf(z)\big|~\big|\vf(z-2\tv')\big|~
         \langle-S\tu'| e^{-tH_0}|S\tu\rangle\,.
\end{equation*}
Integration over $\tu$ and $\tu'$ is now possible,
\begin{equation*}
   \int \frac{d^2\!\tu}{(1+|\tu|^2)^k} = \frac{\pi}{k-1}
   \qquad {\rm and} \qquad
    \int\! d^2\!\tu' ~\langle-S\tu'| e^{-tH_0}|S\tu\rangle
      = \frac{e^{-tm^2}}{4\pi t}\,.
\end{equation*}
As concerns integration over $z$ and $\tv'$, we observe that,
performing the change $\,\tv'\to \half(\tz-\tv')$,
\begin{equation*}
    \int \frac{d^4\!z}{\th(\bz)}~d^2\!\tv'~
       \big|\vf(z)\big|~\big|\vf(z-2\tv')\big|  \,\leq\,
           \|\vf\|_1~\sup_{\bz\in\R^2} \left\{
             \frac{1}{4\,\th^2(\bz)}
                \int d^2\!\tv'~ \big|\vf(\bz,\tv')\big|\right\} \,.
\end{equation*}
All in all we have that $|X_4|\leq{\rm const.}\,e^{-tm^2}/t$, as we
wanted to prove.


\begin{thebibliography}{30}

\bibitem{SeibergW}
N. Seiberg and E. Witten,
``String theory and noncommutative geometry'',
JHEP {\bf 9909} (1999)~032.

\bibitem{Moyal}
J. E. Moyal,
``Quantum mechanics as a statistical theory'',
Proc. Camb. Philos. Soc. {\bf 45} (1949)~99.

\bibitem{Himalia}
V. Gayral, J. M. Gracia-Bond\'{\i}a, B.~Iochum, T.~Sch\"ucker
and~J.~C.~V\'arilly,
``Moyal planes are spectral triples'',
Commun. Math. Phys. {\bf264} (2004) 569.

\bibitem{ConnesGrav}
A. Connes,
``Gravity coupled with matter and the foundation of noncommutative
geometry'',
Commun. Math. Phys. {\bf 182} (1996) 155.

\bibitem{Polaris}
J.~M.~Gracia-Bond\'{\i}a, J.~C.~V\'arilly and H.~Figueroa,
\emph{Elements of Noncommutative Geometry},
Birkh\"auser, Boston, 2001.

\bibitem{GomisM}
J. Gomis and T. Mehen,
``Space-time noncommutative field theories and unitarity'',
Nucl. Phys. {\bf B591} (2000) 265.

\bibitem{GGBRR}
V. Gayral, J. M. Gracia-Bond\'{\i}a and F. Ruiz-Ruiz,
``Trouble with space-like noncommutative field theory'',
Phys. Lett. {\bf B610} (2005) 141.

\bibitem{Bahns}
D. Bahns, S. Doplicher, K. Fredenhagen and G. Piacitelli,
``Field theory on noncommutative spacetimes: quasiplanar Wick products'',
Phys. Rev. {\bf D71} (2005) 025022.

\bibitem{Bahnsetalt}
D. Bahns, S. Doplicher, K. Fredenhagen and G. Piacitelli, 
``On the unitary problem in space-time noncommutative theories'',
Phys. Lett. {\bf B533} (2002) 178.

\bibitem{Fujikawa}
K. Fujikawa,
``Path integral for space-time noncommutative field theory'',
Phys. Rev. {\bf D70} (2004)~085006.

\bibitem{tachyon}
K. Landsteiner, E. L\'opez, M. H. G. Tytgat,
``Excitations in hot noncommutative theories'',
JHEP {\bf 0009} (2000) 027. \\
%
C. P. Mart\'{\i}n and F. Ruiz Ruiz,
``Paramagnetic dominance, the sign of the beta function and UV/IR
mixing in noncommutative $U(1)$'',
Nucl. Phys. {\bf B597} (2001) 197.

\bibitem{FRRsusy}
F. Ruiz Ruiz,
``Gauge-fixing independence of IR divergences in noncommutative U(1),
perturbative tachyonic instabilities and supersymmetry'',
Phys. Lett. {\bf B502} (2001) 274.

\bibitem{Minwallaetal}
S.~Minwalla, M.~V.~Raamsdonk and N.~Seiberg,
``Noncommutative perturbative dynamics'',
JHEP {\bf 0002} (2000) 020.

\bibitem{Matusis}
A. Matusis, L. Susskind and N. Toumbas,
``The IR/UV connection in the noncommutative gauge theories'',
JHEP {\bf 0012} (2000) 002.

\bibitem{Selene}
J. M. Gracia-Bond\'{\i}a, F.~Lizzi, G. Marmo and P. Vitale,
``Infinitely many star-products to play with'',
JHEP {\bf 0204} (2002) 026.

\bibitem{Lizzi}
F. Lizzi, G. Mangano, G. Miele and M. Peloso,
``Cosmological perturbations and short distance physics from
noncommutative geometry'',
JHEP {\bf 0206} (2002) 049.

\bibitem{Dolan}
L. Dolan and C.~R.~Nappi,
``Noncommutativity in a time-dependent background'',
Phys. Lett. {\bf B551} (2003) 369.

\bibitem{CalmetW}
X. Calmet and M. Wohlgennant,
``Effective field theories on noncommutative space-time'',
Phys. Rev. {\bf D68} (2003) 025016.

\bibitem{BehrS}
W. Behr and A. Sykora,
``Construction of gauge theories on curved noncommutative space-time''
Nucl. Phys. {\bf B698} (2004) 473.

\bibitem{Hashimoto}
A. Hashimoto and K. Thomas,
``Dualities, twists and gauge theories with non-constant
noncommutativity'',
JHEP {\bf 0501} (2005) 033.

\bibitem{Cornalba}
L. Cornalba and R. Schiappa,
``Nonassociative star product deformations for D-brane worldvolumes in
curved backgrounds'',
Commun. Math. Phys. {\bf225} (2002) 33.

\bibitem{RieffelDefQ}
M. A. Rieffel,
\emph{Deformation Quantization for Actions of $\R^d$},
Memoirs of the Amer. Math. Soc. {\bf 506}, Providence, RI, 1993.

\bibitem{Gopakumar}
R. Gopakumar, S. Minwalla and A. Strominger,
``Noncommutative solitons'', 
JHEP {\bf 0005} (2000) 020.

\bibitem{covariant}
A.~O.~Barvinsky and G.~A.~Vilkovisky,
``Covariant perturbation theory (II) Second order in the curvature.
General algorithms'',
Nucl. Phys. {\bf B333} (1990) 471.

\bibitem{Sch--DeWitt}
P. B. Gilkey,
\emph{Invariance theory, the heat equation and the Atiyah-Singer
theorem},
CRC Press, Boca Raton, 1995.

\bibitem{Kon}
M. Kontsevich,
``Deformation quantization of Poisson manifolds'',
Lett. Math. Phys. {\bf 66} (2003) 157.

\bibitem{FoscoT}
C.~D.~Fosco and G.~Torroba,
``Noncommutative theories and general coordinate transformations'',
Phys. Rev. {\bf D71} (2005) 065012. 

\bibitem{Groenewold}
H. J. Groenewold,
``On the principles of elementary quantum mechanics'',
Physica {\bf 12} (1946), 405--460.

\bibitem{BMoyal}
M. S. Bartlett and J. E. Moyal,
``The exact transition probabilities of quantum-mechanical oscillators
calculated by the phase-space method'',
Proc. Cambridge Philos. Soc. {\bf 45} (1949), 545--553.

\bibitem{PhobosandDeimos}
J. M. Gracia-Bond\'{\i}a and J. C. V\'arilly,
``Algebras of distributions suitable for phase-space quantum
mechanics. I'',
J. Math. Phys. {\bf 29} (1988) 869.\\
%
J. C. V\'arilly and J. M. Gracia-Bond\'{\i}a,
``Algebras of distributions suitable for phase-space quantum
mechanics. II. Topologies on the Moyal algebra'',
ibidem. {\bf 29} (1988) 880.

\bibitem{Gayral}
V. Gayral,
``Heat-Kernel Approach to UV/IR Mixing on Isospectral Deformation
Manifolds'',
to appear in Ann. Henri Poincar\'e [arXiv:hep-th/0412233].

\bibitem{Himaliatwo}
V. Gayral, B. Iochum and J. C. V\'arilly,
``Dixmier Traces on noncompact isospectral deformations'',
forthcoming.

\bibitem{ConnesLa}
A. Connes and G. Landi,
``Noncommutative manifolds, the instanton algebra and isospectral
deformations'',
Commun. Math. Phys. {\bf 221} (2001) 141.

\bibitem{ConnesDV}
A. Connes and M. Dubois-Violette,
``Noncommutative finite-dimensional manifolds. I. Spherical manifolds
and related examples'',
Commun. Math. Phys. {\bf 230} (2002) 539.

\bibitem{CM}
A. Connes and H. Moscovici,
``Hopf algebras, cyclic cohomology and the transverse index theorem'',
Commun. Math. Phys. {\bf198} (1998) 198.

\bibitem{IZ}
C. Itzykson and J. B.~Zuber, {\it Quantum field theory}, McGraw-Hill,
New York, 1980.

\bibitem{GI}
V. Gayral and B. Iochum,
``The spectral action for Moyal planes'',
to appear in J. Math. Phys. [arXiv:hep-th/0402147].

\bibitem{Araki}
H. Araki,
``Expansional in Banach algebras'',
Ann. Scient. \'Ec. Norm. Sup. {\bf6} (1973) 67.

\bibitem{ConnesBook}
A. Connes,
\emph{Noncommutative Geometry},
Academic Press, London and San Diego, 1994.

\bibitem{FRR}
F. Ruiz Ruiz,
``UV/IR mixing and the Goldstone theorem in noncommutative field
theory'',
Nucl. Phys. {\bf B637} (2002) 143.

\bibitem{Oeckl}
R. Oeckl, 
``Untwisting noncommutative $\R^d$ and the equivalence of
quantum field theories'',
Nucl. Phys. {\bf B581} (2000) 559.\\
%
M. Chaichian, P. P. Kulish, K. Nishijima and A. Tureanu,
``On a Lorentz-invariant interpretation of noncommutative space-time
and its implications on noncommutative quantum field theory'',
Phys.  Lett. {\bf B604} (2004) 98.


\end{thebibliography}
\end{document}